\documentstyle[12pt]{article}

\catcode`\@=11

\def\title{%
\vspace{0.5cm}\vspace{4ex}
\bgroup
\obeylines
\large\boldmath \bf\begin{center}
}
\def\endtitle{\end{center}\vskip1sp\egroup}
\def\author#1{\begingroup\center #1 \endcenter\endgroup}

\def\addcontentsline#1#2#3{\relax}

\def\fnum@figure{Figure \thefigure}
\def\fnum@table{Table \thetable}
\newcounter{figcaption}
\def\thefigcaption{\arabic{figcaption}}
\def\fnum@figcaption{{\bf Fig. \thefigcaption :}}
\def\figcaption{%
{\parindent 0pt \bf Figure Captions} \par \vskip 10pt
\list{\fnum@figcaption}
{\leftmargin 5em \labelwidth\leftmargin\advance\labelwidth-\labelsep
\def\makelabel##1{##1\hfil} \usecounter{figcaption}}%
}

\long\def\@makecaption#1#2{
\vskip 10pt
\begingroup \small
\begin{quote}%
{\bf #1:} #2%
\end{quote}%
\endgroup
}

\def\references#1{\section*{References\@mkboth
{REFERENCES}{REFERENCES}}\list
{ \arabic{enumi})\ }{\settowidth\labelwidth{#1)\ }\leftmargin\labelwidth
\advance\leftmargin\labelsep \usecounter{enumi}}
\def\newblock{\hskip .11em plus .33em minus -.07em}
\sloppy \sfcode`\.=1000\relax}

\def\cite{\@ifnextchar [{\@tempswatrue\@citex}{\@tempswafalse\@citex[]}}
\def\@citex[#1]#2{\if@filesw\immediate\write\@auxout{\string\citation{#2}}\fi
\def\@citea{}\@cite{\@for\@citeb:=#2\do
{\if-\@citeb \mbox{-}\def\@citea{}
\else
\@citea\def\@citea{,\penalty\@m}\@ifundefined
{b@\@citeb}{{\bf ?}\@warning
{Citation `\@citeb' on page \thepage \space undefined}}%
\hbox{\csname b@\@citeb\endcsname}
\fi}}{#1}}
\newfont{\scrptrm}{cmr8}
\def\@cite#1#2{${}^{\scrptrm {#1\if@tempswa , #2\fi})}$}

\def\rcite{\@ifnextchar [{\@tempswatrue\@rcitex}{\@tempswafalse\@rcitex[]}}
\def\@rcitex[#1]#2{\if@filesw\immediate\write\@auxout{\string\citation{#2}}\fi
\def\@rcitea{}\@rcite{\@for\@rciteb:=#2\do
{\if-\@rciteb \mbox{-}\def\@rcitea{}%
\else
\@rcitea\def\@rcitea{,\penalty\@m}\@ifundefined
{b@\@rciteb}{{\bf ?}\@warning
{Citation `\@rciteb' on page \thepage \space undefined}}%
\hbox{\csname b@\@rciteb\endcsname}%
\fi}}{#1}}
\def\@rcite#1#2{{#1\if@tempswa , #2\fi}}
\def\refcite{\@ifnextchar [{\@tempswatrue\@refcitex}
{\@tempswafalse\@refcitex[]}}
\def\@refcitex[#1]#2{
\if@filesw\immediate\write\@auxout{\string\citation{#2}}\fi
\def\@citea{}\@refcite{\@for\@citeb:=#2\do
{\if-\@citeb -\def\@citea{}
\else
\@citea\def\@citea{,\penalty\@m}\@ifundefined
{b@\@citeb}{{\bf ?}\@warning
{Citation `\@citeb' on page \thepage \space undefined}}%
\hbox{\csname b@\@citeb\endcsname}
\fi}}{#1}}
\def\@refcite#1#2{[{#1\if@tempswa , #2\fi}]}

\newcommand{\rmc}{{\rm c}}
\newcommand{\rmd}{{\rm d}}
\newcommand{\rme}{{\rm e}}
\newcommand{\rmi}{{\rm i}}

\renewcommand{\Im}{{\cal I}\!{\sl m}\,}

\newcommand{\Ham}{{\cal H}}

\newcommand{\dags}{^\dagger}

\def\braket#1{\left\langle#1\right\rangle}
\def\brakets#1{\langle#1\rangle}

\def\simle{\mathrel{\mathpalette\@versim<}}   
\def\simge{\mathrel{\mathpalette\@versim>}}   
\def\@versim#1#2{\lower2.5pt\vbox{\baselineskip0pt \lineskip-.5pt
   \ialign{$\m@th#1\hfil##\hfil$\crcr#2\crcr\sim\crcr}}}

\newcommand{\bequ}{ \begin{equation} }
\newcommand{\eequ}{ \end{equation} }
\newcommand{\barr}{ \begin{array} }
\newcommand{\earr}{ \end{array} }
\newcommand{\beqarr}{ \begin{eqnarray} }
\newcommand{\eeqarr}{ \end{eqnarray} }

\newcommand{\baralpha}{ \begin{eqnal} \beqarr}
\newcommand{\earalpha}{ \eeqarr \end{eqnal}}

\def\MFreq{\rmi \omega_n}

\def\vecn{\vec m}
\def\eignmat{I}

\def\LSMO{La$_{1-x}$Sr$_x$MnO$_3$}

\catcode`\@=12

\begin{document}

\begin{title}
Transport Properties of the Kondo Lattice Model
in the Limit  $S=\infty$ and $D=\infty$
\end{title}

\author{Nobuo {\sc Furukawa}}
\begin{instit}
  Institute for Solid State Physics,\\
  University of Tokyo, Roppongi 7-22-1,\\
  Minato-ku, Tokyo 106
\end{instit}


\begin{abstract}
The Kondo lattice model with Hund's ferromagnetic spin
coupling is investigated
as a microscopic model of the perovskite-type 3d transition-metal oxide
{\LSMO}.
In the classical spin limit $S=\infty$ and the
infinite-dimensional limit $D=\infty$,
the one-body Green's function is calculated exactly.
Transport properties of the system in the
presence of   magnetic fields are
calculated.
The giant magnetoresistance of this model,
which is in a good agreement with the experimental data of {\LSMO},
is explained by the spin disorder scattering process.
\end{abstract}

\vfil


\pagebreak

After the discovery of high $T_\rmc$ oxides,
the physics of strongly correlated systems have been
reinvestigated extensively.
One of the main subjects of theoretical and experimental physicists
is the unusual metallic state when the Mott insulator is doped with
carriers.
{}From such motivation,
properties of
3d transition-metal oxide compounds $R_{1-x}A_xM$O$_3$ with
three-dimensional perovskite-type structure
have been intensively studied.
Here $R$ and $A$ denote a trivalent rare-earth ion and a divalent
alkaline-earth ion, respectively, and $M$ represents a 3d transition-metal
 ion.  They are
suited to investigate systematic variation of 3d electron filling,
since the carrier doping
does not destroy the network of transition-metal and oxygen.

Recently,
transport properties for filling-controlled
single crystals of {\LSMO} have been
investigated systematically.\cite{Tokura9x}\ \
For  moderately
doped samples  $x \simge 0.17$, there exists ferromagnetic metal state
below $T_{\rm c}$
due to double exchange mechanism,\cite{deGennes60}\
while at $x\sim 0$ antiferromagnetic insulator phase is observed.
As a model Hamiltonian of this system, the Kondo lattice model (KLM)
in three dimension
\bequ
 \Ham = - \sum_{ij,\sigma} t_{ij}
        \left( c_{i\sigma}\dags c_{j\sigma} + h.c. \right)
    -J \sum_i \vec \sigma_i \cdot \vec S_i
\eequ
with $S=3/2$ and  the Hund's ferromagnetic
 coupling $J>0$ and the strong coupling limit $J \gg |t|$ has been
proposed.\cite{Kubo72}\ \
Here, the notations are standard.
Within this model,
t$_{\rm 2g}$ and  e$_{\rm g}$ electrons of Mn are considered to form
 localized  spins and itinerant fermions, respectively.
The carrier number of the band electron for
{\LSMO} is considered to be $n=1-x$.

At moderately doped region $x = 0.17\sim 0.30$,\cite{Tokura9x}\
a sharp drop is
observed in the resistivity below the magnetic transition
temperature $T_{\rm c}$.
It is also reported that the field induced magnetic moment
also increases the conductivity. Namely, the giant magnetoresistance (GMR)
with negative sign is observed in this material.
In single-crystal samples
of {\LSMO} at $x= 0.15 \sim 0.30$,
the ratios of the magnetoresistance are as large as
 $-\Delta\rho / \rho_0 > 90\%$  around $T_{\rm c}$ $(150\sim350 {\rm K})$.
What is universally seen is that
the resistivity normalized by its zero field value
is  scaled  by a function of the magnetization which has little temperature
 dependence at $T \simge T_{\rm c}$.
In the small magnetization region $M/M_{\rm sat} \simle 0.2$,
the scaling function is given by
\bequ
  \frac{\rho}{\rho_0} = 1 - C \left(\frac{M}{M_{\rm sat}}\right)^2,
    \label{expUnivFit}
\eequ
where $\rho$ is the resistivity and $\rho_0$ is its zero field value.
Here $M_{\rm sat}$ is the saturation magnetization.
The experimental data show that
 the coefficient is $C \sim 4$ at $x \sim 0.17$, 
and the value of $C$ decreases  as the hole concentration $x$
is increased.\cite{Tokura9x,Tokura9xa}\ \

The origin of the resistivity is  qualitatively explained by the
carrier scattering  due to  the thermally fluctuating spin
configurations, or
the spin disorder scattering.
As the spontaneous or the induced magnetic moment is developed,
the amplitude of the  spin fluctuation decreases so that
the resistivity also decreases.
However, existing theories have not succeeded in explaining the
value of $C$ in eq.~(\ref{expUnivFit})
as well as its band filling dependence.
Kubo and Ohata\cite{Kubo72}
 have studied the GMR of this system
using the  effective Hamiltonian.
The resistivity is roughly calculated from the Drude formula with the
quasi-particle life time being estimated from the
thermal fluctuations of spins.
The result at $M \ll M_{\rm sat}$ is described as
\bequ
  \frac{\rho}{\rho_0} = 1 -
   \frac{9}{10} \frac{S(2S+1)}{(S+1)^2}
             \left(\frac{M}{M_{\rm sat}}\right)^2,
\eequ
where $M_{\rm sat} = S$, so that at $S=2$ it
 gives $C = 1$ irrespective of $J$ and $n$.
In the weak coupling approach,
 the Born approximation\cite{Kasuya56} gives
a similar result
\bequ
  \frac{\rho}{\rho_0} = 1- \frac{M(M+1)}{S(S+1)}.
\eequ

In this paper, we study the thermodynamical properties
 of the KLM
with ferromagnetic coupling.
One of our aim is to calculate the conductivity
in order to investigate the GMR of this model and to compare the
coefficient $C$ in eq.~(\ref{expUnivFit}) quantitatively
with the experimental data in {\LSMO}.

Since we consider the case where
 the localized spin is in a high-spin state with the ferromagnetic coupling,
the effect of the quantum exchange
seems to be irrelevant in the paramagnetic phase.
We will later discuss on this point in  detail.
Thus we consider the classical rotator limit $S=\infty$.
The Hamiltonian is described as
\bequ
  \Ham_{S=\infty} =
  - \sum_{ij,\sigma} t_{ij}
        \left(  c_{i\sigma}\dags c_{j\sigma} + h.c. \right)
    -J \sum_i \vec \sigma_i \cdot \vec m_i,
    \label{HamSinfty}
\eequ
where $ \vec m_i = (m_i{}^x, m_i{}^y, m_i{}^z)$ and $|\vec m|^2 = 1$.

Although the Hamiltonian (\ref{HamSinfty})
is only constructed from one-body terms now,
thermodynamic
properties of this system
 is still not easy to obtain.
Here we furthermore take
the limit of infinite dimension $D=\infty$.\cite{Metzner89}\ \
This corresponds to the neglection of
 site off-diagonal terms in  self-energies
and vertex corrections in a realistic $D=3$ lattice system.
In this limit, the problem is
reduced to the single-site problem coupled with a
 self-consistent dynamical
field.\cite{Georges92}\ \

The action of the corresponding single-site model has the form
\beqarr
  S(\tilde G_0,\vecn) &=&
  - \int_0^\beta \rmd \tau_1
   \int_0^\beta \rmd \tau_2\
     \Psi^\dagger (\tau_1) \tilde G_0^{-1}(\tau_1 - \tau_2) \Psi(\tau_2)
        \nonumber \\
  & &  -J  \int_0^\beta \rmd \tau \
      \vecn \cdot \Psi^\dagger(\tau)
           \vec \sigma  \Psi(\tau),
\eeqarr
where $\Psi^\dagger = ( c_\uparrow\dags, c_\downarrow\dags)$ is
the fermion creation operator and $|\vecn| = 1$.
Here, we express the Weiss field  $\tilde G_0 $ by $2\times2$ matrix.
We determine $\tilde G_0$ self-consistently as in ref.~\rcite{Georges92}.
The partition function can be evaluated as
\bequ
  Z = \int \rmd\Omega_{\vecn} Z_{\rm f}(\vecn),
\eequ
where the trace 
$Z_{\rm f}(\vecn)= {\rm Tr} \exp[-S(\tilde G_0,\vecn)] $
 is evaluated as
\beqarr
  Z_{\rm f}(\vecn)
  &=& 4 \exp \left( \sum_n \log \det \left[
       \frac1{\MFreq} (\tilde G_0^{-1} + J\vecn\vec\sigma)
    \right] \rme^{\MFreq 0_{+}} \right).
\eeqarr
The Green's function of the single-site model is calculated exactly as
\bequ
  \tilde G(\MFreq) =
  \braket{ \left(  \tilde G_0^{-1}(\MFreq) +
            J \vecn \vec \sigma \right)^{-1}}_{\vecn},
\eequ
where $\braket{\cdots}_{\vecn}$ represents the thermal average over
$\vecn$ using the ``Boltzman weight'' $ Z_{\rm f}(\vecn) / Z$.

If we restrict ourselves to the paramagnetic solution,
the Green's function is represented by a scalar function as
$  \tilde G_0(\MFreq) =  \tilde g_0(\MFreq) \eignmat$
where $\eignmat$ is the unit matrix, so that
\bequ
  \tilde G(\MFreq) =
    \braket{
      \frac{
            \tilde g_0(\MFreq)^{-1} \eignmat - J \vecn \vec \sigma
      }
      {  \tilde g_0(\MFreq)^{-2} - J^2 |\vecn|^2 }
    }_{\vecn}
     = \frac{  \tilde g_0(\MFreq)^{-1} }
             {  \tilde g_0(\MFreq)^{-2} - J^2}
      \eignmat .
   \label{GFunAve}     \label{GFunSolution}
\eequ
The self-energy is then given by
\bequ
  \Sigma(\MFreq) =
    \tilde G_0^{-1}(\MFreq) - \tilde G^{-1}(\MFreq)
   = J^2 \tilde G_0(\MFreq).
    \label{SelfEnergy}
\eequ
{}From the derivation, we see that the Green's function
(\ref{GFunSolution}) is the same as that of
the system with the Ising substrate spin
$\vecn = (0,0,\pm1)$.

Furthermore,
the Green's function of the
infinite-dimensional Falicov-Kimball model (FKM)
\bequ
  \Ham_{\rm FKM} =
  - \sum_{ij} t_{ij} \left( c_{i}\dags c_{j} + h.c. \right)
    + U \sum_i c_i\dags c_i f_i\dags f_i
\eequ
is described as\cite{Brandt89,Si92}
\bequ
  \tilde G_{\rm FKM}(\MFreq)
    =    \frac{1-\brakets{n_{\rm f}}}{ \tilde G_0^{-1}(\MFreq)}
      +  \frac{\brakets{n_{\rm f}}}{ \tilde G_0^{-1}(\MFreq) - U}.
        \label{GFunFK}
\eequ
Then, the  Green's functions (\ref{GFunSolution}) and (\ref{GFunFK})
share the same analytical structure at $\brakets{n_{\rm f}}=1/2$.
In the FKM, the c-electrons are scattered by the
charge fluctuations of the localized
f-electrons,
which corresponds to the
scattering process of the itinerant electrons by the localized
spins in the KLM.

Therefore, thermodynamical properties
of the KLM in the limit $D=\infty$ and $S=\infty$
can be understood from the nature of
the FKM in infinite dimension which
has been studied intensively.
In the paramagnetic phase,
the imaginary part of the self-energy is finite at the
fermi level,\cite{Si92,Moller92}\
$  \Im \Sigma(0) \ne 0$,
since the spin disorder scattering process remains finite.
Thus the system is not a Fermi liquid.
For $J \gg W$, the density of states splits into lower and upper band
as seen in the Hubbard approximation.
As the magnetic moment is induced, the imaginary part of the
self-energy decreases
because the thermal fluctuation of spins decreases.

Let us now discuss the transport properties of this model
in the presence of a magnetic field.
Calculations are performed using the Lorentzian density of states
for the unperturbed system
\bequ
  N_0(\varepsilon) = \frac1{\pi}  \frac{W}{\varepsilon^2 + W^2},
\eequ
with the band width $W\equiv 1$ as a unit of energy.
Magnetic field is applied to the paramagnetic state.
We express the induced magnetization by $M = \brakets{m_i{}^z}$,
so that $M_{\rm sat} = 1$.
{}From the feature of the Lorentzian density of states,
the Weiss field $\tilde G_0$ is given by
\bequ
  \tilde G_0(\MFreq) =
    \frac{1}{ \MFreq - \mu + \rmi W {\rm sgn}(\omega_n)}
        I,
\eequ
so that the Green's function $\tilde G$ is obtained analytically as
\bequ
  \tilde G = \frac1{ \tilde G_0^{-2} - J^2}
   (\tilde G_0^{-1}I - J M \sigma^z).
   \label{GfunLol}
\eequ
We see that $\tilde G$ as well as
$\tilde\Sigma = \tilde G_0^{-1} - \tilde G^{-1}$ is a function of $M$.
Here, we have
applied magnetic field  only to the localized spin for simplicity.
Effect of the magnetic field applied to electrons
does not make essential difference to the result.

Using the Kubo formula,  optical conductivity is
calculated from the equation\cite{Moller92,Pruschke93a}
\beqarr
  \sigma(\omega) &=& \sigma_0
  \sum_\sigma \int \rmd \omega' \int \rmd \epsilon \
     W^2 N_0(\epsilon) \nonumber \\
  & & \quad \times \
      A_\sigma(\epsilon,\omega') A_\sigma(\epsilon,\omega'+\omega)
   \frac{ f(\omega') - f(\omega'+\omega)}{\omega},
   \label{Optcond}
\eeqarr
where $ A_\sigma(\epsilon,\omega') = - (1/\pi )
        \Im G_{\sigma}(\epsilon,\omega'+ \rmi\eta)$
is the spectral function and $f(\omega)$ is the Fermi distribution
function.
Here we have used the approximation
\bequ
  \frac1{N} \sum_k
   v_{k}^2 \delta( \epsilon - \epsilon_k)
   \simeq \brakets{v^2} \frac1{N} \sum_k
      \delta( \epsilon - \epsilon_k)
   = \brakets{v^2} N_0(\epsilon),
\eequ
and the constant $\brakets{v^2}$ is taken into $\sigma_0$.
This treatment is shown to be exact in the infinite-dimensional
 hypercubic lattice.\cite{Pruschke93a}\ \
The constant $\sigma_0\sim ({ e^2 a^2}/{\hbar}) \cdot ({N}/{V})$
gives the unit of the conductivity
where $a$ is the lattice constant.

{}From eq.~(\ref{Optcond}),
we calculate  the dc conductivity $\sigma_{\rm dc}$.
Since we are interested in the temperature region
$ T_{\rm c} < T \ll \Im\Sigma$ at $J\gg W$,
we may replace $f(\omega)$ in eq.~(\ref{Optcond}) by a step function.
We note that the quasi-particle life time is finite even at the fermi-level
so that $\sigma(\omega)$ at $\omega\to0$ is not singular.
As a typical example, we see $\sigma_{\rm dc} \sim 0.02 \sigma_0$ at
$J=4$ and $\braket{n} = 0.8$ in the paramagnetic state.
The experimental value is
$\sigma_{\rm dc} = 10^1 \sim 10^2 (\Omega {\rm cm})^{-1}$ at $x=0.2$
and $T \simge T_{\rm c}$,
which can be roughly explained by our calculation
since $\sigma_0$ in
three dimension approximately gives Mott's minimum conductivity
 $\sigma_0 \simeq \sigma_{\rm Mott} \sim 10^3(\Omega {\rm cm})^{-1}$.
In Fig.~\ref{FigRhoRho0}, we show the resistivity $\rho$ scaled by its
zero field value $\rho_0$ as a function of $M^2$.
At $M\ll 1$, we clearly see
that $\rho/\rho_0 = 1 - C M^2$ is satisfied.
As shown in ref.~\rcite{Tokura9x}
 for another band filling this curve
reproduces the experimental data very well.
In Fig.~\ref{FigRhoCoeff}, we show the coefficient $C$ as a function of
$J$ and $n$.
In the weak coupling limit $J \ll 1$ we have $C=1$
so that the
result of Born approximation at $S=\infty$
is reproduced.
 We see that as $J$ is increased,
a deviation from the weak coupling limit is observed.
The value of $C$ at the strong coupling region $J \gg W$
decreases monotonically as the system is doped with holes.
Thus,
the experimental value of $C$ at lightly doped region as well as
its doping dependence can be explained
from the current model in the strong coupling region.

In the paramagnetic phase and
the weakly spin polarized state 
of {\LSMO},
we see that $\sigma_{\rm dc} \ll \sigma_{\rm Mott}$ so that
the system is in a very incoherent state,
and the model calculation also shows that
the quasi-particle excitation is incoherent.
The above results imply that the system has a short coherence length.
Therefore, the effectively single-site treatment of the $D=\infty$ system
is justified.
The time scale of the quantum spin-flip process at $J \gg W$
is estimated by $1/T_{\rm K} \sim 1/W$ which is relatively longer than the
quasi-particle life time $1/\Im\Sigma \sim W/J^2$.
Then, it is also justified
to take the limit of  $S=\infty$.
However, for the coherent metal state in
the low temperature ferromagnetic phase of
{\LSMO},
the limit of $S=\infty$ and $D=\infty$ may not be justified.
Quantum mechanical treatment of the localized spins as well as
the spatially collective excitations may be necessary for the calculation.

To summarize, we have calculated the KLM in
the limit $S=\infty$ and $D=\infty$. The Green's function
is obtained exactly,
which appears to have the same analytical structure as that of the
Falicov-Kimball model.
In the paramagnetic phase,
the quasi-particle excitation
has the finite life time due to the spin disorder scattering.
Using the Kubo formula,
we have investigated the transport properties of the model
under the magnetic field.
As also shown in ref.~\rcite{Tokura9x},
the GMR in {\LSMO} is quantitatively reproduced by the present calculation.

The author would like to thank Y. Tokura, A. Urushibara and
F.~F. Assaad
for fruitful discussions and comments.
He is grateful to
Y. Kuramoto and T. Kawarabayashi for interesting discussions
at the early stage of this work.

\begin{figcaption}

\item Magnetization dependences of $\rho/\rho_0$ at $\braket{n}=0.8$.
The result of the Born approximation (BA) is also shown.
\label{FigRhoRho0}

\item Coefficient $C$ for various $J$ and $n$.
\label{FigRhoCoeff}
\end{figcaption}

\end{document}